# Scanning probe imaging of coexistent ferromagnetism and ferroelectricity at room temperature

V. R. Palkar, S. C. Purandare, S. Gohil, J. John and S. Bhattacharya

*Department of Condensed Matter Physics and Materials Science, Tata Institute of Fundamental Research, Mumbai 400005, India.*

Abstract

Room temperature coexistence of ferromagnetism and ferroelectricity in a thin film of a novel material of nominal composition $PbTi_{0.5}Fe_{0.5}O_{3-\delta}$ is probed by standard ferroelectric and ferromagnetic hysteresis loop measurements and by scanning probe microscopy of various kinds. Both magnetic domains and ferroelectric domains are observed in the same spatial region of the material, implying phase coexistence in this system. For both order parameters, sample morphology strongly affects roughness of the domain walls.

______________________________________________________________________

Recent resurgence of interest in multiferroic systems in general, and in magnetoelectric systems in particular, is driven by two different reasons, namely, the basic physics of coexistence of two order parameters in the ground state and the possibility of using this coexistence, and a possible coupling between them, for devising novel applications [1-2]. Magnetoelectric systems, which exhibit coexistence of magnetism and ferroelectricity with a possible coupling between the two order parameters, are of particular importance, especially when the magnetism is of the ferromagnetic variety. Multifunctional and exotic non-volatile memory applications are possible in that case. However, the presence of local spins and of off-centered structural distortion are basic requirements for magnetic and ferroelectric properties, respectively and naturally, these two seemingly unrelated phenomena rarely co-exist at room temperature. Unfortunately, most of the known multiferroic systems are antiferromegnetic with transition temperatures below room temperature [3,4]. Therefore, for device applications these materials hardly find any use. Recently, we have reported novel materials exhibiting magnetoelectric properties at room temperature [5-7]. In this letter we report room temperature multiferroic properties for PbTi$_{0.5}$Fe$_{0.5}$O$_{3-\delta}$ thin films in macroscopic scale from hysteresis measurements. Moreover, simultaneous application of scanning probe microscopy techniques, viz., Magnetic Force Microscopy (MFM) [8] for ferromagnetism and Piezo-respsonse Force Microscopy (PFM) and Electric Force Microscopy (EFM) for ferroelectricity [9], have revealed small scale coexistence of both ferromagnetic and ferroelectric domains in the same spatial region.

It was shown recently [7] that introduction of Fe in PbTiO$_3$, a conventional ferroelectric material, induces ferromagnetism without disturbing its ferroelectric nature. It was feasible due to the presence of large lattice distortion in the parent compound (c/a ~ 1.066). As expected, substitution with Fe (transition metal) at Ti site brings reduction in c/a ratio. However, though suppressed, the lattice still remains distorted enough even for the composition PbTi$_{0.5}$Fe$_{0.5}$O$_{3-\delta}$ (c/a ~ 1.025). Thus it is possible to convert ferroelectric lead titanate into multiferroic system. Significantly, both ferroelectric (~420 ºC) and ferromagnetic transitions (~280 ºC) occur above room temperature in bulk samples. It may be noted that the magnetic transition temperatures for Fe, Fe$_2$O$_3$ and Fe$_3$O$_4$ are 770, 643 and 575 ºC respectively and the magnetic transition temperature observed for the compound with nominal composition PbTi$_{0.5}$Fe$_{0.5}$O$_{3-\delta}$ is much lower compared to any of these phases. It is essential to reproduce these properties in thin films for their usefulness in practical applications, as well as for further basic physics investigations. In this letter we report that thin films of this material, grown on Pt/TiO$_2$/SiO$_2$/Si substrate by using

Pulsed Laser Deposition (PLD) technique, preserve the room temperature multiferroic nature of this system. More importantly we are successful in demonstrating the phase coexistence occurring at microscopic scale as well.

In order to achieve good chemical homogeneity and controlled stoichiometry, $Pb(Ti_{0.5}Fe_{0.5})O_{3-\delta}$ powder sample was prepared by using wet chemical route (co-precipitation). The detailed procedure for synthesis is described elsewhere [7]. From Nominal composition (with $Fe^{+3}$ replacing $Ti^{+4}$) one would expect the system to be an electrical conductor. Instead, we have been able to synthesize samples, which are insulating. Powder samples thus prepared were compacted and sintered at 1000 °C/2hr. to obtain dense target for the laser ablation process. Sintering was carried out in PbO atmosphere to avoid PbO losses. The densely compacted target helped to minimize particulate growth commonly occurring during ablation process and thus improved the film quality. The pulsed laser deposition process used for growing thin films has been described earlier [10]. Deposition conditions such as substrate temperature, substrate to target distance, laser energy and oxygen pressure were optimized so as to achieve single phase, highly insulating thin films (resistivity of the order of $10^8$-$10^9$ $\Omega$.cm), similar to that of target material. It was observed that oxygen pressure in the chamber during ablation process plays very crucial role in controlling film quality. The phase purity was confirmed by X-ray diffraction. Figure 1 (a) shows the X-ray diffraction pattern of the thin films compared with that for the bulk material. The films were polycrystalline in nature without any preferential orientation. The ferroelectric order parameter, is directly proportional to the tetragonal lattice distortion (c/a). The lattice distortion is about 1.025, significantly reduced from the value of 1.066 in the parent compound ($PbTiO_3$) implying a reduction due to the Fe-substitution in the lattice due to formation of solid solution. The lattice parameter values of the film match with that of bulk target. Suppression in lattice distortion has reflected on ferroelectric transition temperature. Differential Scanning Calorimetric studies on films indicated two peaks (change in slope) around 260 ºC and 410 ºC, similar to that of bulk material (Fig. 1b). The first peak is attributed to that of magnetic transition while second peak indicates ferroelectric transition [7]. The films were characterized by Scanning Electron Microscopy (SEM) and Energy Dispersive X-ray Analysis (EDX). It was seen that the films were granular in nature and the matrix was uniform. EDX studies revealed that the films were chemically homogeneous and no segregation of any phase/element in the matrix was observed. X-ray mapping study also supported the EDX results.

In Fig 1(c) is shown the magnetic M-H hystersis loop obtained at room temperature by a Quantum Design SQUID Magnetometer. The magnetization saturates around an applied field of 1 Tesla, although the average coercive magnetic field is less than 1 kGauss. The saturation magnetization is as high as 74emu/cc which corresponds to a magnetic moment of ~1 Bohr magneton per Fe atom, significantly smaller than the moment in Iron or Iron oxides. Fig. 1 (d) shows the ferroelectric hysteresis loop of the thin film, obtained by using a ferroelectric loop tracer from Radiant Technology. The saturation occurs around an applied field of 200kV/cm; the coercive field is ~ 66kV/cm. Both the saturation polarization value of 30uC/cm$^2$ and the retentivity of 60% as measured by the ratio of the remnant polarization to the saturated value are typical of standard ferroelectric materials attesting to its usefulness in memory application. In other words, the system displays reduced ferroelectricity (value for single crystal $PbTiO_3$ ~ 80 uC/cm$^2$) and reduced ferromagnetism (2.2 $u_B$ for pure Fe) but still retains enough magnitude of both order parameters for being practically useful.

These results establish the multi-ferroic nature of the thin films. However, both from the basic physics questions as well as exotic applications, one requires a direct demonstration of the phase co-existence of magnetic and ferroelectric domain structure at microscopic level; in other words it is most critical to establish that both order parameters are not microscopically phase-separated in different spatial regions of the sample. This can be studied by scanning probe microscopy and we use a Multimode Atomic Force Microscope (Nanoscope IV from Digital Instruments) configured to provide, simultaneously, both the magnetic and the ferroelectric domain structure, in a specific spatial area of the thin film spanning tens of microns on the lateral scale with a spatial resolution better than 50 nanometer.

For the magnetic force microscopy (MFM), a tapping cantilever with a tip coated with cobalt was used. The tip was magnetized with a strong permanent magnet before installing the tip holder on the AFM head. Tapping mode topographic image was obtained during the main scan in order to correlate the nature of the local order parameter with sample morphology. During the interleave mode, the tip was raised above the sample surface allowing the imaging of relatively weak but long-range magnetic interactions while minimizing the influence of topography. The resulting images from the raw data for topography and magnetic structure are shown in fig. 2(a) without any additional image processing. "Up" and "down" domains in the form of light and dark stripes, commonly observed in ferromagnetic films, are seen.

Magnetic domain patterns are in general determined by the thermal history of the film. Magneto-crystalline crystal structure and grain morphology are the two dominating factors that influence the domain formation. Generally it is expected that each grain in a polycrystalline sample should have a magneto-crystalline structure typical of large single crystal. However, there is always modification due to the interaction with neighbouring grains. Therefore in case of thin films, though magnetocrystalline anisotropy dominates the domain formation, grain morphology becomes deciding factor with increase in thickness. Magnetocrystalline anisotropy is an intrinsic property independent of grain size and shape. Anisotropic energy directs magnetization along certain crystallographic axis called direction of easy magnetization. Magnetic anisotropy strongly affects the shape of the hysteresis loops and controls the coercivity and remanence. It could be easily detected by measuring magnetization curves along different crystal direction. Depending upon the crystallographic orientation of the sample in the magnetic field, magnetization reaches saturation at different fields. $PbTi_{0.5}Fe_{0.5}O_{3-\delta}$ with pseudo-cubic crystal structure is likely to have [100] directions as easy direction of magnetization and [111] directions are hard direction. SQUID measurements (M-H curve) carried out on our films in the range of thickness ~200nm and above in parallel and perpendicular geometries are almost similar. It perhaps suggests that magnetocrystallinity is not a factor as far as domain configuration is concerned in these films but is mainly dominated by grain morphology.

Variation of the domains from zigzag to stripes to bubble pattern with film thickness was observed. This could be due to the fact that the grain size changed with thickness of the film. It will be reported elsewhere in detail. We note, however, that the graininess of the domain walls corresponds well with the grain structure although the domain walls maintain large scale persistence extending across many grains. A detailed understanding of the competing effects of grain morphology and the intrinsic domain structure will be important for both, an understanding of the basic question of the role of quenched disorder on the correlation length of the order parameter as well as for the design of a possible exotic memory element.

Fig. 2(b) shows the simultaneous measurement of the topography with the piezoresponse measurement. The same tip is used in this case with an external electronic circuitry involving an external dc field in order to polarize the film and an ac field to detect the piezoresponse, applied between the metallic substrate below the film and the metallic tip. Piezoresponse Force microscopy (PFM) technique is used in which the amplitude and phase of the piezoresponse, related to the integrated amplitude and direction of the polarization, are imaged. For simplicity we show simultaneous measurement of the topography and phase alone. The sample is first rastered with a large negative voltage to polarize the entire area in the up direction, seen as dark background. Then a positive switching voltage is applied to "write" a square in the middle, which is seen as light colour. The polarizing voltage is turned off and the remnant state is then imaged, by using the ac field alone. Comparing the topography images, we ascertain that the same spatial area is indeed scanned in this case. The remnant state is clearly seen. Once again, the graininess of the topography is visible in the remnant domain walls implying strong pinning of the domain walls on the grain boundaries, although the overall poled region extends across many grains. This result demonstrates the following. (1) The same area, measuring 20 microns on the side shows simultaneous existence of both magnetic stripe domains as well as the written, albeit imperfect, square of the ferroelectric domain. In other words, the two order parameters are coexistent not only in the macro-scale but also in the microscopic scale. (2) In both cases, the grain morphology modulates the micro scale domain structure. The absence of visible domain walls in the intragranular space suggests that the grains themselves prefer to retain a single value of polarization and magnetization. While this observation is common in ferroelectric materials, it is less so for conventional metallic ferromagnets.

It now remains to be seen what effects the ferroelectric polarizing and writing of a domain have on the previously observed magnetic domain structure. During the piezoresponse mode, the entire 20μm region is first polarized with positive dc voltage. The center 10μm square region is then oppositely polarized with negative dc voltage. This is shown in Fig. 2(b). Fig. 2(c) shows MFM image taken after piezoresponse imaging, that is, image taken in tapping mode after writing the square as shown in Fig. 2(b). Remarkably, while the topography is unaltered to the accuracy of the measurement, the MFM image appears to be a composite of the previous two images. The stripe domains are still present as in Fig. 2(a), but so is the remnant polarization as in Fig. 2(b). Fig. 2(c) shows magnetic domains and written ferroelectric domains superimposed on each other and the contrast due to magnetic domains in the form of stripes superimposed on ferroelectrically polarized dark region is clearly evident. The same magnetic stripe domains, although present in the central bright region, where the electric polarization is reversed as compared to the outer dark region, are not so easily discernable since both the stripes and the ferroelectric domain give rise to bright contrast. If the outer region was polarized with negative voltage, and the central region was polarized with positive voltage, then the stripes would be clearly visible in the central region but not so in the outer region. Though these are MFM measurements, what we observe is the effect of both magnetic force and electric force due to remnant polarization. In other words, this is a simultaneous MFM and EFM image. *We also note that the stripes, both dark and bright in (a) have become darker in the dark region of (b) while the same stripes in (a) have become brighter in the bright region of (b).* Thus both forces are contributing to the measured force as is indeed expected from the technique itself. This is further confirmed by repeating the same set of measurements in pure lead titanate films. In that case we observe the same effect as in fig 2 with the exception of the magnetic stripes, as expected from the absence of magnetism in lead titanate. This confirms the composite nature of the image as seen in 2(c). We are not aware of a prior observation of this kind in any condensed matter system.

In view of the composite nature of the image, a direct determination of the coupling between the two order parameters from the imaging study becomes far more complex and requires a detailed deconvolution and absolute calibration, which is beyond the scope of this study. But the images themselves attest to the feasibility of such a direct evaluation, at least in principle.

We have thus shown room temperature multiferroic behavior of $Pb(Ti_{0.5}Fe_{0.5})O_{3-\delta}$ thin films at the microscopic level. Remarkably, imaging of magnetic domains and reversal of ferroelectric domain switching in the same spatial area of the film, using several scanning probe imaging techniques, demonstrates the coexistence of the two parameters at microscopic level also.

In conclusion, the coexistence of ferroelectric and magnetic ordering at room temperature particularly at microscopic level has been observed in thin films for the first time. The composite image indicating presence of both order parameters at microscopic level is remarkable. The results reported here are not only clear and unambiguous, but reproducible also. From Nominal composition (with $Fe^{+3}$ replacing $Ti^{+4}$) one would expect the system to be an electrical conductor. Instead, we have been able to synthesize samples, which are insulating with resistivity as high as $10^8$-$10^9$ $\Omega$.cm. The compensation of oxygen charge imbalance remains a mystery. A detailed analysis of the microstructure, small-scale stoichiometry and physical properties are under way in order to understand this critical issue. Neutron diffraction study is one of this kinds. We acknowledge helpful discussions with G. Ravikumar and B. A. Chalke.

References

[1]. M. Fiebig, J. Phys. D: Appl. Phys. 38, R123 (2005)

[2]. W. Prellier, M. P. Singh and P. Murugavel, J. Phys.: Condens. Matter 17, R803 (2005)

[3]. N. A. Hill, J. Phys. Chem. B 104, 6694 (2000)

[4]. N. A. Spaldin & M. Fiebig, Science, 309, 391 (2005)

[5] V. R. Palkar, Darshan Kundaliya, S. K. Malik & S. Bhattacharya, Physical Review B 69, 212102 (2004)

[6]. V. R. Palkar, K. Ganesh Kumara, & S. K. Malik, Applied Physics Letters 84, 2856 (2004)

[7] V. R. Palkar & S. K. Malik, Solid State Communications, 134(11), 783 (2005)

[8].Han, B. S. Physics, 26(10), 617-624 (1997)

[9] Nanoscale phenomena in Ferroelectric thin films, Edited by S. Hong (Kluwer Academic, Boston, 2004)

[10]. V. R. Palkar, S. C. Purandare, S. P. Pai, S. Chattopadhyay., P. R. Apte, R. Pinto & Multani M. S. Appl. Phys. Letter 68 (11), (1996)

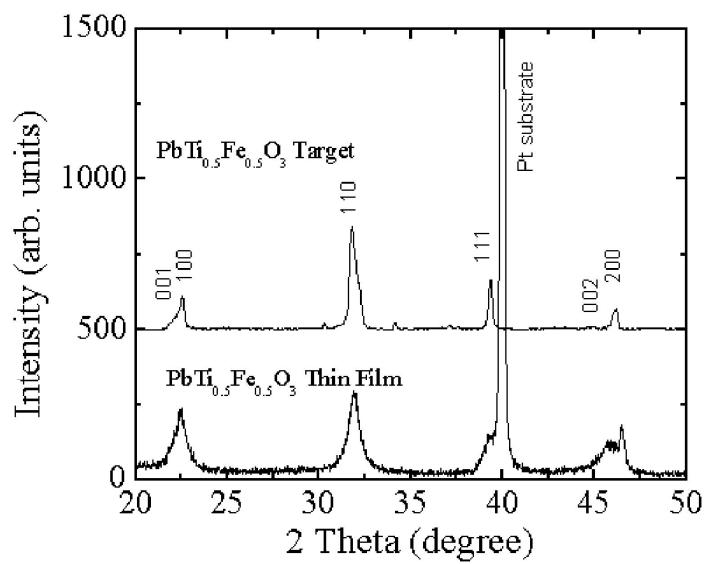

Figure 1(a). XRD pattern obtained for PbTi$_{0.5}$Fe$_{0.5}$O$_{3-\delta}$ target and thin film (~3000 Å) grown on Pt/TiO2/SiO2/Si substrate using Pulse Laser Deposition technique. The typical conditions used for deposition are, laser energy – 250 mj and frequency – 15Hz, target to substrate distance – 4.5 cm, oxygen partial pressure –500 mtorr, substrate temperature – 650 ºC and deposition time-30 min. The film is single phase with tetragonal crystal structure. (lattice parameter *a*=b=3.98 Å, parameter *c*-4.08 Å and *c/a*-1.025).

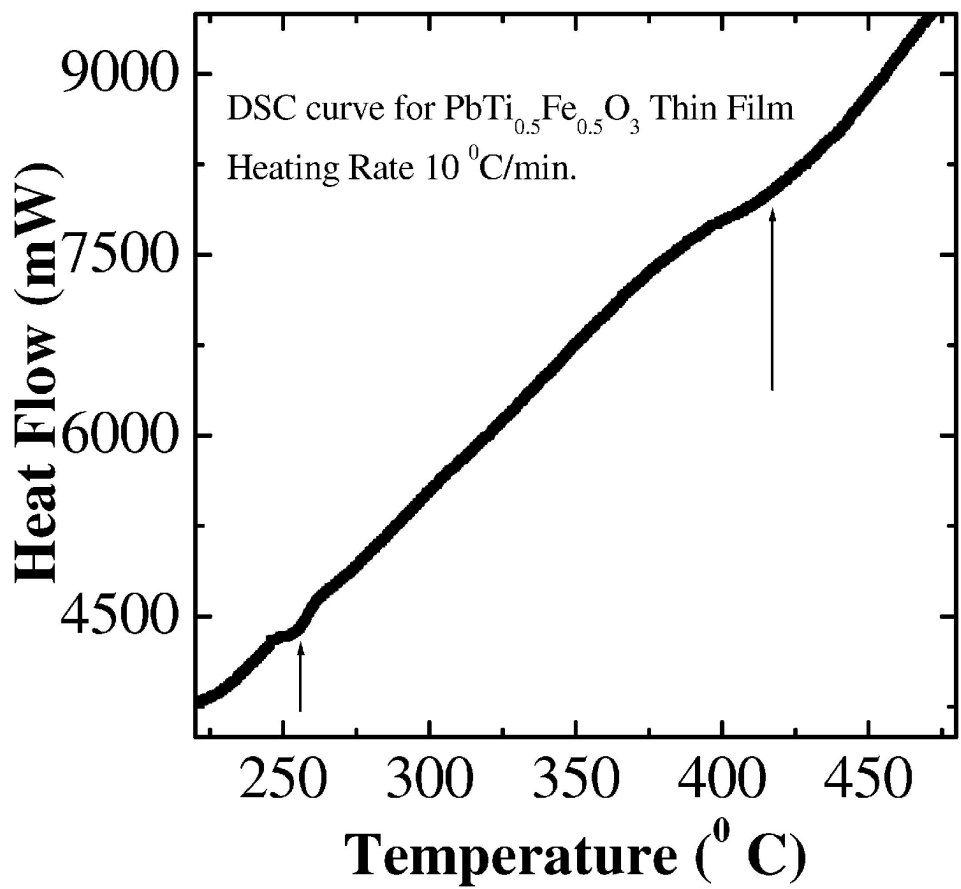

Figure 1(b). Differential Scanning Calorimetry curve for $PbTi_{0.5}Fe_{0.5}O_{3-\delta}$ thin film. Heating rate 10 °C/min.

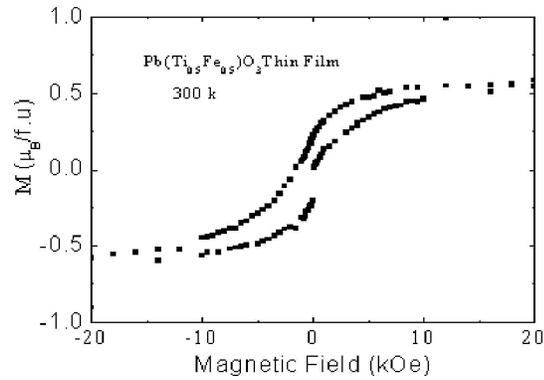

Figure 1(c). M-H isotherm obtained at room temperature using a Quantum Design SQUID magnetometer for PbTi$_{0.5}$Fe$_{0.5}$O$_{3-\delta}$ thin film.

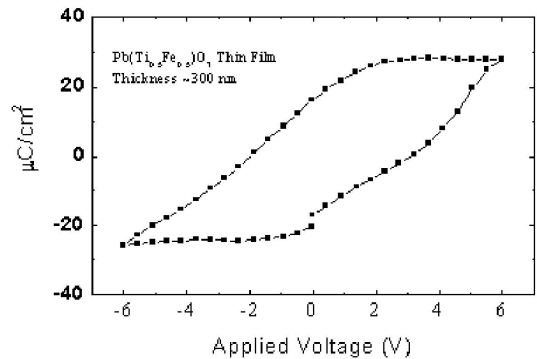

Figure 1(d). Ferroelectric loop measured by using RT66A ferroelectric loop tracer on PbTi$_{0.5}$Fe$_{0.5}$O$_{3-\delta}$ thin film (~3000 Å). Pt is used as top and bottom electrode.

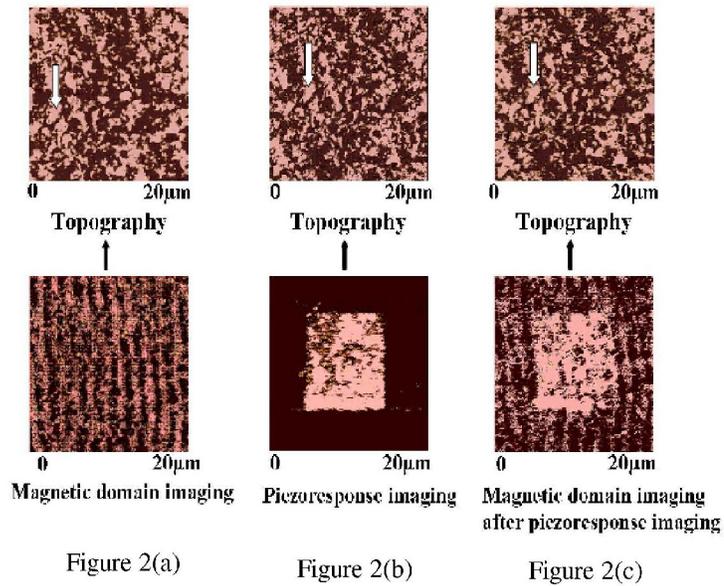

Figure 2. Multimode scanning probe images of PbTi$_{0.5}$Fe$_{0.5}$O$_{3-\delta}$ thin film. In all three panels, the topography was measured simultaneously (shown on the top panel in each case) to ascertain that the same spatial region is imaged. A small positional shift could not be avoided and this shift is ascertained by marking the corresponding grains with arrows.

(a). Topography and magnetic domain imaging of PbTi$_{0.5}$Fe$_{0.5}$O$_{3-\delta}$ thin film using Magnetic Force Microscopy. The magnetic stripe domains consisting of up and down magnetization, typical of ferromagnetic thin films, are observed. Typical spacing between stripes is about 1micron. The roughness of the domain structure corresponds well with the grain structure but each stripe extends across many grains.

(b). Topography and piezoresponse imaging of PbTi$_{0.5}$Fe$_{0.5}$O$_{3-\delta}$ thin film using Electric Force Microscopy. The image shows unambiguously the ferroelectric writing performed by the poling voltages and the existence of remanent polarization after the poling voltages were switched off. Note that stripes are not observed in this image, as expected from the imaging technique.

(c). Topography and domain imaging of $PbTi_{0.5}Fe_{0.5}O_{3-\delta}$ thin film using Magnetic Force Microscopy in tapping mode after writing ferroelectric domains (2b). This tapping mode image appears to contain information of both ferromagnetic and ferroelectric domain structure and is a composite of the two previous images. See text for discussions.